\documentclass[%
jcp,
superscriptaddress,
 amsmath,amssymb,
 preprint,%
 ]{revtex4-1}
\usepackage{graphicx}
\usepackage{color}
\usepackage{dcolumn}
\usepackage{physics}
\usepackage{algorithm}
\usepackage{algpseudocode}
\usepackage{bm}
\usepackage{epstopdf}
\usepackage{makecell}

\usepackage[utf8]{inputenc}
\usepackage[T1]{fontenc}
\usepackage{mathptmx}
\usepackage{xcolor}
\global\long\def\tr{\mathrm{Tr}}

\newcommand{\veck}{\mathbf{k}}

\begin{document}


\title[Projected Density Matrix Embedding Theory]{Projected Density Matrix Embedding Theory with
Applications to the Two-Dimensional Hubbard Model}

\author{Xiaojie Wu}
\thanks{Equal contributions}
\affiliation{Department of Mathematics, University of California, Berkeley, CA 94720}

\author{Zhi-Hao Cui}
\thanks{Equal contributions}
\affiliation{Division of Chemistry and Chemical Engineering, California Institute of Technology, Pasadena, CA 91125}

\author{Yu Tong}
\affiliation{Department of Mathematics, University of California, Berkeley, CA 94720}

\author{Michael Lindsey}
\affiliation{Department of Mathematics, University of California, Berkeley, CA 94720}

\author{Garnet Kin-Lic Chan}
\email{gkc1000@gmail.com}
\affiliation{Division of Chemistry and Chemical Engineering, California Institute of Technology, Pasadena, CA 91125}

\author{Lin Lin}
\email{linlin@math.berkeley.edu}
\affiliation{Department of Mathematics, University of California, Berkeley, CA 94720}
\affiliation{Computational Research Division, Lawrence Berkeley National Laboratory, Berkeley, CA 94720}


\begin{abstract}
Density matrix embedding theory (DMET) is a quantum
embedding theory for strongly correlated systems. From a
computational perspective, one bottleneck in DMET is the optimization of the correlation potential to achieve self-consistency, especially
for heterogeneous systems of large size. We propose a new method,
called projected density matrix embedding theory (p-DMET), which 
achieves self-consistency without  needing to optimize a
correlation potential. We demonstrate the performance of p-DMET on the two-dimensional Hubbard model. 
\end{abstract}

\maketitle

\section{Introduction}\label{sec:intro}

Strong correlation effects play an important role in many quantum
systems and require treatment beyond the level of mean-field
theories. However, the absence of a mean-field starting
point means that a direct treatment of strong correlations,
for example, by full configuration interaction (FCI)~\cite{knowles1984new,olsen1990passing,vogiatzis2017pushing} or
exact diagonalization (ED)~\cite{lin1990exact,lauchli2011ground}, scales exponentially with respect to system size.
This motivates the development of
quantum embedding theories~\cite{sun2016quantum,imada2010electronic,ayral2017dynamical}, which partition the global system
into a series of fictitious and strongly correlated fixed-size ``impurities'', which are then 
treated accurately via a high-level theory (such as FCI or ED). The solutions from all impurities are then coupled
together via a lower-level theory. This procedure often needs to be performed
self-consistently. Examples of such quantum embedding theories include 
dynamical mean field theory (DMFT)~\cite{metzner1989correlated, georges1992numerical,georges1996dynamical,kotliar2006electronic} and density matrix
embedding theory (DMET)~\cite{DMET2012, DMET2013, tsuchimochi2015density, bulik2014density, wsj16}. DMET has been successfully applied to
compute phase diagrams of a number of strongly correlated models, such as the one-band Hubbard model on different lattices~\cite{DMET2012, bulik2014density, chen2014intermediate, boxiao2016, Zheng2017, zheng2017stripe}, quantum spin models~\cite{Fan15, gunst2017block}, and some
prototypical correlated molecular problems~\cite{DMET2013, wsj16, Pham18}.


This work focuses on improving numerical algorithms to
achieve self-consistency in the context of DMET. In DMET, each
impurity problem consists of two parts: a fragment and a bath. 
The low-level theory is used to construct the bath, which is defined by the Schmidt decomposition of the low-level wavefunction between the fragment and the remaining
part of the global system.  The standard choice in DMET is to choose the
 low-level theory to be a mean-field theory
(such as Hartree-Fock theory) so that the low-level wavefunction is a Slater
 determinant. In this case, the corresponding one-particle reduced density matrix
 (1-RDM, also simply referred to as the density matrix here) is idempotent, and
 the bath space is fully spanned by a set of one-particle bath orbitals,
 which can be efficiently obtained by factorizing the 1-RDM between the one-particle degrees of freedom
associated with the fragment and those in the global
system. The high-level theory is then used to evaluate the 1-RDM and
the two-particle reduced density matrix (2-RDM) associated with each
impurity, which can be assembled according
to the DMET democratic partitioning protocol~\cite{wsj16} to evaluate the total energy and other physical observables, such as correlation functions.


In many systems, using the bath orbitals generated from
the Hartree-Fock Slater determinant yields energies and physical observables 
from DMET that already significantly improve on those from Hartree-Fock theory alone. 
Such calculations will be referred to as ``single-shot'' DMET calculations~\cite{wsj16}. On the other hand, when the physical system undergoes a phase transition not predicted by mean-field theory, we expect that a mean-field theory will produce the wrong order parameter, and the resulting bath orbitals
will be very poor. In such a scenario, it is necessary to perform DMET \textit{self-consistently} to improve the bath orbitals. The self-consistency condition is usually defined such that the 1-RDMs
obtained from the low-level and high-level theories match each
other according to some criterion, such as matching the density matrix in the impurity problem ~\cite{DMET2012}, on the fragment only~\cite{DMET2013,tsuchimochi2015density}, or simply matching the diagonal elements of the density matrix (i.e. the electron density)~\cite{bulik2014density}.  Self-consistency can be achieved  by optimizing a single-body potential, termed the correlation potential, in the low-level theory. 
Each optimization step requires diagonalizing a matrix, similar to in a self-consistent field (SCF) iteration step in the solution of the Hartree-Fock equations.


Nonetheless, there are two outstanding numerical issues associated with
the optimization of the correlation potential. First,  the optimization
procedure may require a large number of iterations to converge. It is
not uncommon for the number of iterations to be $100\sim 1000$
especially for systems that are not translationally invariant. Hence when
the system size becomes moderately large (a few hundred sites), the cost
of the correlation potential optimization may 
exceed the cost of the impurity solver for small impurities. Second, the bath construction
procedure of DMET requires the 1-RDM to be an idempotent matrix, and the
corresponding low-level Hamiltonian should have a finite HOMO-LUMO gap.
However, even if the strongly correlated global system is gapped, it is
often the case that the low-level Hamiltonian associated with a given
correlation potential in the optimization procedure becomes gapless.
The derivative of the bath orbitals with respect to the correlation
potential will then become infinite, and the optimization cannot properly
proceed. This work aims at addressing the first problem, namely the cost associated with the correlation potential optimization.
The second problem should be addressed by properly considering the zero temperature limit of a finite temperature generalization of DMET,
which will be studied in the future.

We will introduce an alternate procedure to self-consistently determine the bath orbitals,
which completely avoids the need to optimize the correlation potential. In the standard DMET, the bath orbitals are
uniquely determined by the corresponding idempotent 1-RDM obtained from a low-level theory, denoted $D^{\text{ll}}$. The goal of the self-consistent DMET
can then be formulated, in an abstract way, as finding the solution of the following fixed point problem 
\begin{equation}\label{eqn:fixedpoint}
D^{\text{ll}}=\mathcal{F}\left[\mathcal{D}[D^{\text{ll}}]\right].
\end{equation}
Here the mapping $\mathcal{D}[\cdot]$ takes the idempotent 1-RDM as
input, generates the corresponding bath orbitals, and solves all impurity
problems to obtain the 1-RDM evaluated from the high-level theory.  The
mapping $\mathcal{F}$ takes the high-level correlated 1-RDM, denoted by
$D^{\text{hl}}:=\mathcal{D}[D^{\text{ll}}]$ as input, and generates
another idempotent 1-RDM. The correlation potential optimization can be
viewed as \textit{one} way of achieving self-consistency as required
by~\eqref{eqn:fixedpoint}. To see this, we only need to define  the
mapping $\mathcal{F}$ to be the minimization procedure in the standard
DMET, which uses a correlation potential to minimize the discrepancy
between $D^{\text{ll}}$ and 
$D^{\text{hl}}$ evaluated on the impurity problems.

The perspective from the fixed point equation~\eqref{eqn:fixedpoint} suggests that other forms of $\mathcal{F}$ are possible which map $D^{\text{hl}}$ to $D^{\text{ll}}$ more efficiently.  We propose that $D^{\text{ll}}$ can be obtained by directly \textit{projecting} $D^{\text{hl}}$ onto the set of idempotent matrices with a given rank. The modified method is therefore called the projected density matrix embedding theory (p-DMET). The solution of p-DMET will \textit{not} be identical to that of DMET, since they are defined using different mappings $\mathcal{F}$. In particular, unlike DMET, which can be defined to only use information from $D^{\text{hl}}$ on the fragments during self-consistency, p-DMET requires the construction of $D^{\text{hl}}$ on the global domain in order to define the projection operation. 
 
Using the two-dimensional one-band Hubbard model and restricting
to magnetic and non-magnetic self-consistent solutions, we demonstrate that the results of p-DMET and DMET at self-consistency
are very similar within the anti-ferromagnetic (AFM) and the paramagnetic (PM) phases. The discrepancy between the two methods is largest near the phase boundary, and for larger on-site interactions.
We show that p-DMET significantly lowers the computational cost to achieve self-consistency for large lattices
  without translational invariance. For example,
even for a moderately sized lattice with 128 sites and using 4 site impurities (without translational invariance), the correlation potential fitting procedure in standard DMET
requires about $20000$ s of CPU time, which is reduced to about $1$ s in the p-DMET approach.
 
The rest of the paper is organized as follows. In Section~\ref{sec:DMET} we
briefly review DMET and the correlation potential optimization problem. We 
then discuss p-DMET and the associated numerical issues in Section~\ref{sec:PDMET}. We demonstrate the performance of p-DMET for the two-dimensional Hubbard model in Section~\ref{sec:numer}, before deriving conclusions in Section~\ref{sec:conclusion}. The procedure for constructing the bath orbitals from the low-level RDM is summarized in Appendix~\ref{sec:bath}. 

\section{Review of density matrix embedding theory}\label{sec:DMET}

\begin{figure}[!htp]
  \centering
  \includegraphics[scale=0.4]{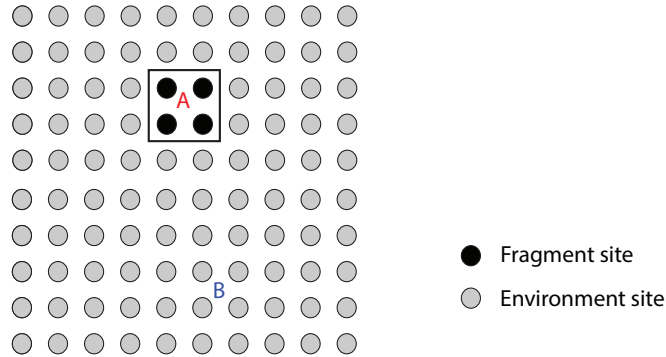}
  \caption{Illustration of a DMET 2-by-2 fragment embedded in a
  two-dimensional square lattice.}
  \label{fig:dmet}
\end{figure}

Let us consider a system of $L$ sites and $N_e$ electrons, which is
partitioned into $N_f$ non-overlapping fragments. Without loss of
generality, we assume the system is spinless; spinful systems can
be represented as spinless systems by doubling the system size.  We
assume
all of the fragments contain $L_A$ lattice sites. 
Fig.~\ref{fig:dmet} shows a $10\times 10$ lattice partitioned into
fragments of size $2\times 2$ ($L_{A}=4$).
The (particle number-conserving) Hamiltonian of the global system takes the form
\begin{equation}
  \label{eqn:ham}
  \hat H = \sum_{pq}^L t_{pq}\hat a_p^\dagger \hat
  a_q+\frac{1}{2}\sum_{pqrs}^L (pr|qs)\hat a_p^\dagger\hat
  a_q^\dagger\hat a_s\hat a_r ,
\end{equation}
where $\hat a_p^\dagger$ ($\hat a_q$) are electron creation (annihilation) operators and $t_{pq}$ and $(pr|qs)$ are one- and two-electron integrals respectively. 

For any given fragment
$A$, the ground state wave function of the global system $\ket{\Psi}$ can always
be partitioned as 
\begin{equation}
  \label{eq:24}
  \ket{\Psi} =
  \sum_{i=1}^{N_A}\sum_{j=1}^{N_B}\Psi_{ij}\ket{A_i}\ket{B_j}.
\end{equation}
Here $\ket{A_i}$ represents a state in the fragment, and $\ket{B_j}$
represents a state in the environment $B$. $N_A$ and $N_B$ are the
numbers of states in the fragment and environment, respectively. In
general, $N_B \gg N_A$.

Conceptually, we may perform a singular value decomposition (also called the Schmidt decomposition in the current context) of the
coefficient matrix $\Psi_{ij}$, which yields
\begin{equation}
  \label{eqn:schmidt}
  \ket{\Psi} =
  \sum_{i=1}^{N_A}\sum_{j=1}^{N_B}\sum_{\alpha=1}^{N_A}U_{\alpha
  i}\lambda_\alpha V^\dagger_{\alpha j}\ket{A_i }\ket{ B_j
  }=\sum_{\alpha=1}^{N_A}\lambda_\alpha \ket{\widetilde A_\alpha
  }\ket{\widetilde B_\alpha}.
\end{equation}
Here $\{\lambda_\alpha\}$ are the singular values. The unitary matrices
$U$ and $V$ are absorbed into the transformed states $\ket{\widetilde
A_\alpha}$ and $\ket{\widetilde B_\alpha }$. $\ket{\widetilde B_\alpha}$ defines
a bath state for fragment $A$. Eq.~\eqref{eqn:schmidt} suggests that the
maximum number of bath states is bounded by $N_{A} \leqslant 2^{L_{A}}$. 

For each fragment, we define the impurity Hilbert space
to be the product space of the 
fragment states $\{\widetilde A_\alpha^{(x)}\}$ and the associated
bath states. Hence a basis for the impurity
$x$ can be written as
$\ket{\widetilde A_\alpha^{(x)} \widetilde
B_\beta^{(x)}}_{\alpha,\beta=1}^{N_{A}}$. We can then construct the projection
operator onto the impurity space as 
\[
\hat{P}^{(x)} = \sum_{\alpha,\beta}\ket{\widetilde A_\alpha^{(x)}
\widetilde B_\beta^{(x)}}\bra{\widetilde A_\alpha^{(x)} \widetilde
B_\beta^{(x)}} .
\]
In the interacting bath formulation of DMET (the only formulation
we explicitly compute with in this work) this projector
conceptually defines each impurity Hamiltonian via
\begin{equation}
  \label{eqn:Hembed}
  \hat H_\text{emb}^{(x)} = \hat{P}^{(x)}\hat H\hat{P}^{(x)}. 
\end{equation}


In most applications of DMET including in this work, we further
require that the global wavefunction used to define the bath space in Eq.~\eqref{eqn:schmidt}
is a Slater determinant. In this case,
the bath states are themselves determinants of two types
of \textit{single particle orbitals}: bath orbitals (usually fractionally occupied in the bath states) and core
orbitals (fully occupied in all bath states). Both the bath and core orbitals can be obtained directly from
the idempotent 1-RDM of the global Slater determinant, and the procedure is summarized in Appendix~\ref{sec:bath}.
Because of this correspondence, the projector $\hat{P}^{(x)}$ can be replaced by a projector onto
the Hilbert space of the fragment and bath orbitals. The contribution of the core orbitals amounts to an embedding potential in $\hat H_\text{emb}^{(x)}$. 
We refer the reader to Ref. \cite{wsj16} for a more detailed description of these steps.


Given the impurity Hamiltonians \eqref{eqn:Hembed} we can  solve for each of the impurity ground-states.
In the special case of translational invariant systems, we can solve for a single impurity ground-state, and
use the fact that expectation values of all other impurities can be obtained by translation.
From 1-RDM's computed from these ground-states we can assemble a high-level correlated 1-RDM $D^{\text{hl}}$ for
the global system according to a chosen partitioning, such as the democratic partitioning~\cite{wsj16} (also see section \ref{sec:PDMET}). Many numerical experiments show that obtaining the correct number of electrons from $D^{\text{hl}}$ is extremely important for the accuracy of observables. However, the number of electrons computed from the high-level correlated 1-RDM $D^{\text{hl}}$, that is, $\Tr(D^{\text{hl}})$, generally does not match the target number of electrons in the system $N_e$. A global chemical potential $\mu$ is often introduced on the fragment part (excluding the bath) of each impurity Hamiltonian. More specifically,
\begin{equation}
    \hat H_\text{emb}^{(x)} \leftarrow \hat H_\text{emb}^{(x)} - \mu
    \begin{bmatrix}
    I_{L_A} & 0\\
    0 & 0
    \end{bmatrix}.
\end{equation}
Here $I_{L_A}$ is an $L_A\times L_A$ identity matrix. The condition $\Tr(D^{\text{hl}})=N_e$ is achieved by adjusting the global chemical potential. The global chemical potential can be viewed as a Lagrangian multiplier to enforce the correct number of electrons in $D^{\mathrm{hl}}$ in the optimization problem.

To formulate DMET in a self-consistent fashion, we then refer to Eq.~\eqref{eqn:fixedpoint}. For a given correlated 1-RDM
$D^{\text{hl}}$, the mapping $\mathcal{F}$ is defined as
\begin{equation}
  D^\text{ll}:=\mathcal{F}[D^{\text{hl}}] = \underset {D\in \mathcal{A}}{\arg \min} 
   \|(D^{\text{hl}}-D)\odot W\|^2_F.
  \label{eqn:Fmap_dmet}
\end{equation}
Here $\odot$ is the element-wise product, and $W$ is a weight matrix. 
If we choose $W$ to be the identity matrix, then
Eq.~\eqref{eqn:Fmap_dmet} only measures the discrepancy of the 1-RDMs
on the diagonal entries. This gives rise to the density embedding
theory~\cite{bulik2014density}. We may also let $W_{pq}=0$ if the indices $p,q$
do not belong to the same fragment, and otherwise $W_{pq}=1$. In this case,
$W$ is a block diagonal matrix and each block is a matrix of ones. Eq.~\eqref{eqn:Fmap_dmet} measures the sum of discrepancies on each fragment~\cite{wsj16}. 

In standard DMET, the admissible set $\mathcal{A}$ is the subset of idempotent
density matrices generated by a correlation potential added to a one-particle Hamiltonian. More
precisely, let $u_{pq}$ be a sparse Hermitian matrix, satisfying the condition $u_{pq}=0$ if $p,q$ do not belong to the same fragment. The effective
one-particle Hamiltonian of the low-level theory is
\begin{equation}
  \label{eqn:meanfield}
  \hat{h}^{\text{ll}} = \hat{a}^{\dagger} (t+u) \hat{a},
\end{equation}
where 
\begin{equation}
\hat{a}^{\dagger} = \begin{bmatrix}  \hat a_1^\dagger& \dots &\hat a_L^\dagger\end{bmatrix}, 
\quad 
\hat{a} = \begin{bmatrix}
    \hat a_1 & \dots & \hat a_L
  \end{bmatrix}^{T}
\end{equation}
define the collection of all creation operators and annihilation
operators, and $u$ is called the correlation potential. $D^{\text{ll}}$ is the ground-state 1-RDM corresponding to
$\hat{h}^{\text{ll}}$ satisfying $\tr[D^{\text{ll}}]=N_{e}$, where $N_e$ is the number of electrons in the system. For this procedure to be well defined, there should be a positive gap between the $N_e$-th and $(N_e+1)$-th eigenvalues of 
$\hat{h}^{\text{ll}}$. 
Under this condition, the optimization problem~\eqref{eqn:Fmap_dmet} can be equivalently formulated as finding
the optimal correlation potential $u$.

When $u$ is a block diagonal Hermitian matrix, the number of
independent variables is $\frac{(L_A+1)L_A}{2}\cdot\frac{L}{L_A} =
\frac{(L_A+1)L}{2}.$ The minimization can be carried out using
derivatives with respect to $u$, which 
can be
explicitly calculated using perturbation theory.
The optimization problem can then be numerically
solved by quasi-Newton or conjugate gradient type methods. As discussed
in Section~\ref{sec:intro}, the optimization of the correlation
potential can be a bottleneck in larger systems, particularly when translational
  invariance is not present.


\section{Projected density matrix embedding theory}\label{sec:PDMET}


Motivated by the fixed point formulation of self-consistency~\eqref{eqn:fixedpoint}, we
propose the following procedure to obtain
$D^{\text{ll}}$:
\begin{equation}
  \label{eqn:pdmet_w}
  D^\text{ll} = \mathcal{F}[D^{\text{hl}}]:=\arg \min_{\substack{D=D^\dagger, D^2=D,\\
  \tr(D)=N_e}}\|(D-D^\text{hl})\odot W\|_F^2.
\end{equation}
Compared to~\eqref{eqn:Fmap_dmet}, the main simplification of
Eq.~\eqref{eqn:pdmet_w} comes from the fact that the admissible set is now the set of idempotent density
matrices with $N_{e}$ electrons without further constraints. To further
simplify the method we let each entry 
of $W$ be $1$, i.e. $W_{pq}\equiv
1$. In other words, we measure the discrepancy of all entries of the
density matrix on the same footing, and solve
\begin{equation}
  \label{eqn:pdmet}
  D^\text{ll} = \arg \min_{\substack{D=D^\dagger, D^2=D,\\
  \tr(D)=N_e}}\|D-D^\text{hl}\|_F^2.
\end{equation}

Eq.~\eqref{eqn:pdmet} has a simple analytic solution. Let
$\Psi^{\text{ll}}$ be the eigenvectors corresponding to the largest 
$N_{e}$ eigenvalues of $D^{\text{hl}}$, i.e. $\Psi^{\text{ll}}$
consists of the leading $N_{e}$ natural orbitals. Then the solution to
Eq.~\eqref{eqn:pdmet} is
\begin{equation}
  \label{eqn:pdmet_sol}
  D^\text{ll} = \Psi^{\text{ll}}(\Psi^{\text{ll}})^\dagger.
\end{equation}
If $D^{\text{hl}}$ is fixed, this is the closest projection operator to $D^\text{hl}$ measured by
the Frobenius norm. Informally, $\Psi^\text{ll}$ is the  single determinant that best captures
the information contained in all the density matrices of the fragments.
Once $D^\text{ll}$ is obtained, we may compute the
bath orbitals and the core orbitals according to the procedure in
Appendix~\ref{sec:bath}, and proceed to solve for the ground-state of the impurity problems as in
the standard DMET procedure. Hence we refer to this method as 
projected density matrix embedding theory (p-DMET). Again to make this procedure well defined, we require that there is a positive gap between the $N_{e}$-th and $(N_{e}+1)$-th eigenvalue of the correlated density matrix $D^{\text{hl}}$. p-DMET assumes the high-level 1-RDM $D^{\text{hl}}$ of the global
system has been computed. As mentioned above, such a global 1-RDM can be constructed
from the high-level 1-RDM's in each impurity using the  democratic partitioning. This procedure was briefly described in~\cite{wsj16}
and we give a more explicit description for completeness now.



\begin{figure}[!htp]
  \centering
  \includegraphics[scale=0.4]{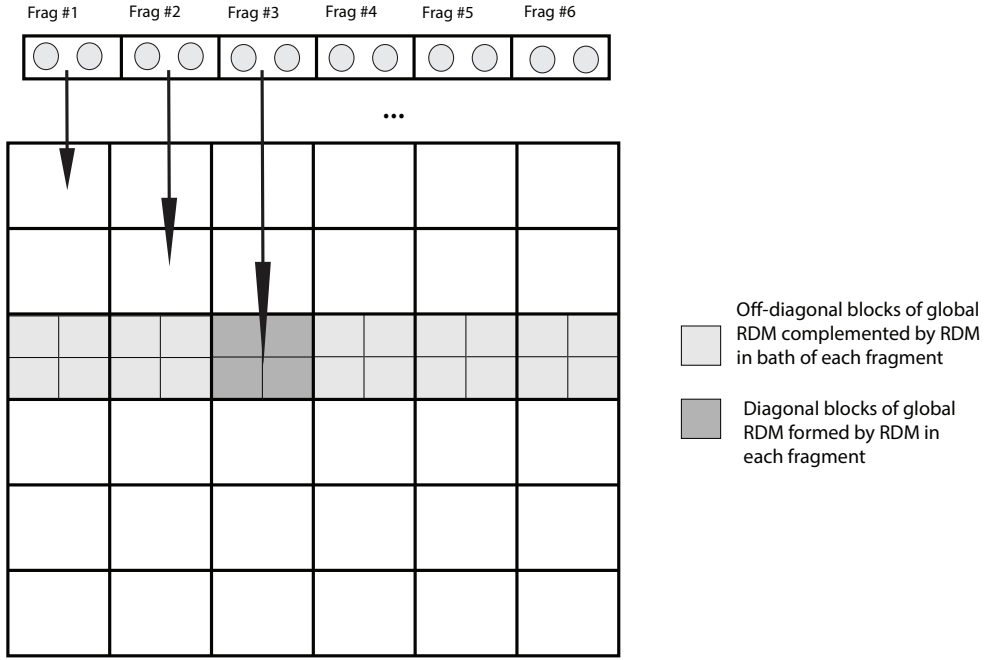}
  \caption{Construction of global density matrix. Each fragment
  contributes a rectangular block row of the global density matrix. Each
  block row has the same length as the global density matrix.}
  \label{fig:glob_dm}
\end{figure}

Let $C_{x}$ be the collection of fragment and bath orbitals of the
impurity $x$ as in Eq.~\eqref{eqn:Cx}.  Then we define
\begin{equation}
  \label{eq:4}
  \begin{bmatrix}
    D^{(x)}\\
    *
  \end{bmatrix}
  =\widetilde D^{(x)}C_x ^\dagger,
\end{equation}
where $\widetilde D^{(x)}$ is the 1-RDM of the impurity problem with
size $(2L_A)\times (2L_A)$.  $\widetilde D^{(x)}C_x^\dagger$ is a matrix
of size $(2L_A)\times L$. $D^{(x)}$ is obtained by extracting the first
$L_A$ rows of $\widetilde D^{(x)}C_x^\dagger$, which is a
block row of the density matrix corresponding to the fragment part of
the impurity $x$. Since the fragments collectively form a non-overlapping
partitioning of the global system, the global 1-RDM can be formed as
\begin{equation}
  \label{eqn:dm_cor}
  D^{\text{hl}} =
  \begin{bmatrix}
    D^{(1)}\\
    D^{(2)}\\
    \vdots\\
    D^{(N_f)}
  \end{bmatrix} .
\end{equation}
Fig. \ref{fig:glob_dm} illustrates the procedure of constructing
$D^{\text{hl}}$ for a one-dimensional model with $12$ sites
partitioned into $6$ fragments.

Since each block row of $D^{\text{hl}}$ is obtained from the
corresponding impurity problem independently, in general
$D^{\text{hl}}$ is not a Hermitian matrix. Therefore, after
Eq.~\eqref{eqn:dm_cor}, we symmetrize the 1-RDM as
\begin{equation}
D^\text{hl}\gets \frac{ D^{\text{hl}}+( D^{\text{hl}})^\dagger}{2}.
\end{equation}
This symmetrization procedure corresponds to the choice of  ``democratic
partitioning'' for constructing the 1-RDM (and its contribution to
the total energy) in DMET~\cite{wsj16}.
In general,  $D^{\text{hl}}$ will not be an idempotent matrix. Hence there
is a non-zero discrepancy between $D^{\text{hl}}$ and $D^{\text{ll}}$.


%
%
%

In order to solve the fixed point problem~\eqref{eqn:fixedpoint} in
p-DMET, 
an extrapolation, or mixing scheme is usually beneficial to accelerate the convergence.
In p-DMET we choose $D^{\text{hl}}$ as the mixing variable. Let
$D^{\text{hl},(k)}$ be the correlated 1-RDM at the beginning of the
$k$-th iteration. Then we first compute the low-level density matrix
$\overline D^{\text{ll}}=\mathcal{F}\left[D^{\text{hl},(k)}\right]$ through the
projection~\eqref{eqn:pdmet_sol}, construct the corresponding bath orbitals, and
solve the impurity problem. From this we obtain an output correlated density
matrix $\overline D^{\text{hl}}:=\mathcal{D}[\overline D^{\text{ll}}]$. Define
the residual as
\begin{equation}
  R^{(k)} := D^{\text{hl},(k)} - \overline D^{\text{hl}}.
  \label{eqn:pdmet_residual}
\end{equation}
The simplest scheme to obtain $D^{\text{hl},(k+1)}$ is the simple mixing
\begin{equation}
   D^{\text{hl},(k+1)} = D^{\text{hl},(k)} - \alpha R^{(k)}.
  \label{eqn:pdmet_simplemix}
\end{equation}
Here $0<\alpha \leqslant 1$ is a mixing parameter.  The simple mixing method
usually converges when $\alpha$ is set to be sufficiently small, but the
convergence rate can be very slow. In order to accelerate the convergence,
we can use the direct inversion in the iterative subspace (DIIS)
method~\cite{Pulay1980}. In DIIS, $D^{\text{hl},(k+1)}$ is obtained by
extrapolating the 1-RDM's from the previous $\ell+1$ steps as
\begin{equation}
  D^{\text{hl},(k+1)} = \sum_{j=k-\ell}^k\alpha_j D^{\text{hl},(j)}.
  \label{}
\end{equation}
In order to obtain the mixing coefficient
$\{\alpha_{j}\}_{j=k-\ell}^{k}$, we also record the residual
$\{R^{(j)}\}_{j=k-\ell}^{k}$ as in Eq.~\eqref{eqn:pdmet_residual}, and solve
the following minimization problem
\begin{equation}
  \label{eq:19}
  \{\alpha_j\} =\underset{\{\alpha_{j}\}} \arg \min \left\|
  \sum_{j=k-\ell}^k\alpha_j R^{(j)}\right\|_F, \text{ s.t. }
  \sum_{j=k-\ell}^k\alpha_j=1.
\end{equation}

A pseudocode implementation for the p-DMET and DMET algorithm is provided in Algorithm~\ref{alg:pdmet}.
\begin{algorithm}[H]
  \caption{A unified pseudocode for the projected density matrix embedding theory (p-DMET) and density matrix embedding theory (DMET).} 
  \label{alg:pdmet}
  \begin{algorithmic}[1]
    \leftline{\textbf{Input:} Initial guess of the correlated 1-RDM $D^{\text{hl},(0)}$ and chemical potential $\mu^{(0)}$.}
    \leftline{\textbf{Output:} Converged correlated 1-RDM $D^\text{hl}$
      and ground state energy $E$.}
    \For{$k = 0, \ldots,$}
    \State $\mu^{(k,0)} \gets \mu^{(k)}$
    \For{$m = 0, \ldots,$} 
    \For{each impurity $x$}
    \State Compute fragment orbitals ($C_f$), and bath orbitals ($C_b$) \Comment{Section: Appendix  \ref{sec:bath}}
    \State Formulate the impurity Hamiltonian, $\hat H_{\text{emb}}^{(x)}$ \Comment{Eq. (16,17) in \cite{wsj16}}
    \State Solve the impurity problem with $\mu^{(k,m)}$ \Comment{via solvers such as FCI or DMRG}
    \EndFor
    \State Compute the total number of electrons, $\tr({D^\text{hl}})$
    \State if convergence is not reached, update $\mu^{(k,m+1)}$ by Newton's iterations
    \EndFor
    \State $\mu^{(k+1)} \gets \mu^{(k,m)}$
    \State Construct the correlated 1-RDM $\overline D^\text{hl}$ \Comment{Eq. \eqref{eqn:dm_cor} followed by symmetrization}
    \State Compute energy $E^{(k)}$ from $\overline D^\text{hl}$ as well as the related 2-RDM \Comment{Eq. (28) in \cite{wsj16}}
    \State If convergence is reached, exit the loop
     \If {embedding method is p-DMET}
    \State \underline{Perform mixing scheme to obtain $D^{\text{hl},(k+1)}$} 
    \State \underline{Compute low-level density matrix $D^\text{ll}$} \Comment{Eq. \eqref{eqn:pdmet_sol}}
    \ElsIf {embedding method is DMET}
    \State \underline{Solve the minimization problem} \Comment{Eq. \eqref{eqn:Fmap_dmet}}
    \State \underline{Perform mixing scheme to obtain a new correlation potential $u^{(k+1)}$}
    \EndIf
    \EndFor
    \State Set $D^{\text{hl}}\gets D^{\text{hl},(k)}$, $E\gets E^{(k)}$
  \end{algorithmic}
\end{algorithm}

In principle, one could also choose the mean field density matrix  $D^{\text{ll}}$ as the mixing variable. However, there arises a practical question related to this choice, namely that the linear combination of a few (or even two) idempotent matrices is generally not an idempotent matrix. Note that the same problem already arises in the context of Hartree-Fock calculations. Some of us have recently developed the projected commutator DIIS (PC-DIIS) method~\cite{hu2017projected}, which accelerates Hartree-Fock calculations within a large basis set (such as planewaves). The idea of PC-DIIS is to apply the idempotent density matrix $D^{\text{ll}}$ to a gauge-fixing matrix $\Phi^{\text{ref}}$ as $\Phi=D^{\text{ll}}\Phi^{\text{ref}}$. It
is clear that the information in $D^{\text{ll}}$ and $\Phi$ is
equivalent. In particular, $D^{\text{ll}}$ can be reconstructed from
$\Phi$ (L\"owdin orthogonalization) as
\begin{equation}
  \label{eqn:dm_reconstruct}
  D^{\text{ll}} = \Phi(\Phi^{\dagger}\Phi)^{-1}\Phi^{\dagger}.
\end{equation}
Thus PC-DIIS uses $\Phi$ as the mixing variable, and reconstructs the
idempotent density matrix using Eq.~\eqref{eqn:dm_reconstruct}. 

Following the PC-DIIS method, we can then choose $\Phi$ as the mixing variable in the self-consistent p-DMET. The gauge-fixing matrix $\Phi^{\text{ref}}$ can
be chosen to be, for instance, the Hartree-Fock occupied orbital coefficient matrix.
We refer to this method as the
projected density matrix embedding theory with a fixed gauge
(p-DMET-f). Note that from the perspective of Eq.~\eqref{eqn:fixedpoint},
p-DMET-f solves the same fixed point problem as p-DMET. The only difference is the choice of the mixing variable.

For translational invariant systems, note that $D^{\mathrm{hl}}$ constructed from democratic partitioning does not break the translational symmetry among fragments. Therefore, the $D^{\mathrm{hl}}$ can be represented in $\veck$-space, denoted by $D^{\mathrm{hl}} (\veck)$. The corresponding $\Psi^{\mathrm{ll}} (\veck)$ is generated per $\veck$ sector. The extrapolation over $D^{\mathrm{hl}} (\veck)$ also conserves the crystal momentum $\veck$. Finally, we note that p-DMET-f can be formulated in a similar way by introducing a gauge-fixing matrix  per $\veck$ sector.

\section{Numerical Experiments}\label{sec:numer}

In this section, we investigate the performance of p-DMET and p-DMET-f
for a 2D Hubbard model with periodic boundary conditions. The mean-field
theory is chosen to be the unrestricted Hartree-Fock (UHF) theory, and the impurity
Hamiltonian in DMET and p-DMET is defined within the interacting bath formulation as in~\eqref{eqn:Hembed}.  The impurity ground-states were computed using the FCI method implemented in the
\textsc{PySCF}~\cite{PYSCF} package and the density matrix renormalization group (DMRG) method~\cite{White92, White93}, as implemented in the \textsc{block} program \cite{Chan02, Chan04, Chan11, Sharma12}, using a bond dimension of $M = 1000$, the split-localized orbital strategy described in~\cite{boxiao2016}, and the genetic algorithm for orbital ordering \cite{Olivares-Amaya15}. 
The fragments were chosen to be $2\times 2$ clusters, treated without translational invariance, to allow a comparison between p-DMET and DMET in
a general setting, except in the case of the cluster size convergence tests, where fragments of up to $4 \times 4$ (16 sites) were used,
and translational invariance was assumed.
We used a convergence criterion on the energy difference between two
consecutive iterations of less than $10^{-8}$ for the $2\times 2$ clusters and $10^{-5}$ for the larger clusters. All energies are reported in units of hopping ($t$).


\subsection{Accuracy}\label{subsec:accuracy}

\begin{figure}[!htp]
  \centering
  \includegraphics[scale=0.5]{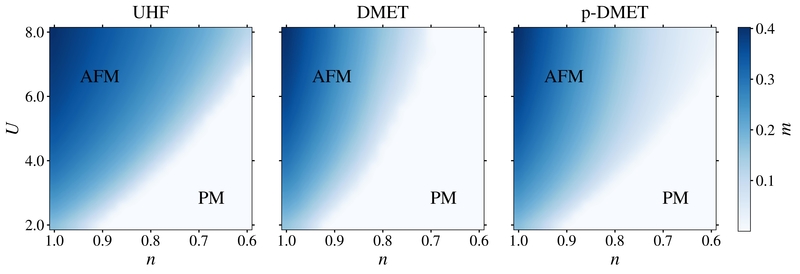}
  \caption[phase diagram]{Phase diagrams of the 2D Hubbard model from UHF, DMET and p-DMET. The color represents the spin polarization ($m=\frac{1}{2}|n_\uparrow-n_\downarrow|$), where $n_\uparrow$ and $n_\downarrow$ are spin up and spin down densities respectively. }
  \label{fig:phase}
\end{figure}

To investigate the accuracy of p-DMET, we plot the phase diagram of a 2D Hubbard model with $40 \times 40$ sites with periodic boundary conditions. This system has been studied in~\cite{boxiao2016} using a translationally-invariant implementation of DMET.  The initial 1-RDM is produced by a converged UHF calculation. Fig. \ref{fig:phase} compares the phase diagrams
generated by UHF, DMET, and p-DMET respectively, evaluated on a
$21\times 21$ grid with respect to the on-site interaction strength
$U$, as well as the filling factor $n$. 

The phase diagram is divided into two regions distinguished by their spin polarization, i.e. the anti-ferromagnetic (AFM)
phase and the paramagnetic (PM) phase. The phase diagrams obtained from p-DMET and DMET are qualitatively similar. The two diagrams agree well with each other when $U \leqslant 4.0$, and larger discrepancies between p-DMET and DMET are observed in the region $U > 4.0$ and $0.6 \leqslant n \leqslant 0.8$. 
We also observe that the the phase boundary obtained from p-DMET is slightly softer, i.e. the decay of the spin polarization from the AFM phase to the PM phase is slower than that in DMET.  

A quantitative comparison of the total energy per site can be found in Table \ref{tab:energy}. Overall, the discrepancy between p-DMET and DMET is much smaller than that between UHF and DMET. The energies of p-DMET and DMET agree very well (the difference is less than $10^{-3}$) inside the AFM / PM phases. The largest discrepancy occurs at $U=6.0,n=0.750$, again near the phase boundary, where the difference of the energy is $0.022$. We remark that neither p-DMET nor DMET is variational, so we cannot determine from this single calculation which is better.


\begin{table}[!htp]
  \centering
    \caption[Energy per site]{Energy per site of the 2D Hubbard model by UHF, p-DMET and DMET  as a function of $U$ and doping ($n$). The self-consistent p-DMET and DMET calculations use the converged UHF solution as the initial guess.  
  }
  \label{tab:energy}
  \begin{tabular}{c|ccc|ccc|ccc}
    \hline
    & \multicolumn{3}{c}{$n=1.000$} & \multicolumn{3}{c}{$n=0.875$} & \multicolumn{3}{c}{$n = 0.750$}\\
    \cline{2-4}  \cline{5-7}  \cline{8-10}
    $U$ & UHF & p-DMET  & DMET  & UHF & p-DMET & DMET & UHF & p-DMET  & DMET \\
    \hline
    $2.0$ & -1.13886 & -1.17999 & -1.17985 & -1.22470 & -1.27817 & -1.27799 & -1.27655 & -1.32270 & -1.32275 \\
    $4.0$ & -0.79703 & -0.86792 & -0.86856 & -0.88440 & -1.03002 & -1.03450 & -0.99530 & -1.16862 & -1.16707 \\
    $6.0$ & -0.59270 & -0.66099 & -0.66188 & -0.66592 & -0.87265 & -0.87395 & -0.75936 & -1.04709 & -1.06860 \\
    $8.0$ & -0.46588 & -0.52262 & -0.52393 & -0.52665 & -0.77299 & -0.77149 & -0.60439 & -0.97734 & -0.98954 \\
    \hline
  \end{tabular}
\end{table}

\subsection{Convergence}\label{subsec:convergence}

We observe in Fig.~\ref{fig:phase} that the softer phase boundary in p-DMET coincides with the region where UHF and DMET predict different phases. Since the UHF solution only enters p-DMET as an initial guess, we may wonder whether the fixed-point of p-DMET  depends on the initial guess. Below we demonstrate that the converged  p-DMET/p-DMET-f solution can indeed depend on the initial guess, at least in certain parts of the phase diagram.  We consider a 2D Hubbard system with $6\times 6$ sites with periodic boundary conditions. The onsite interaction $U$ is set to $4.0$, and we consider two fillings:  $n=1.0$ (half filling, $N_e=36$), and $n=0.722$ ($N_e=26$). In both cases, the energy gap at the mean-field level is positive, and the self-consistent procedure for all methods are well defined without any finite temperature smearing.

\begin{figure}[!htp]
  \includegraphics[scale=0.7]{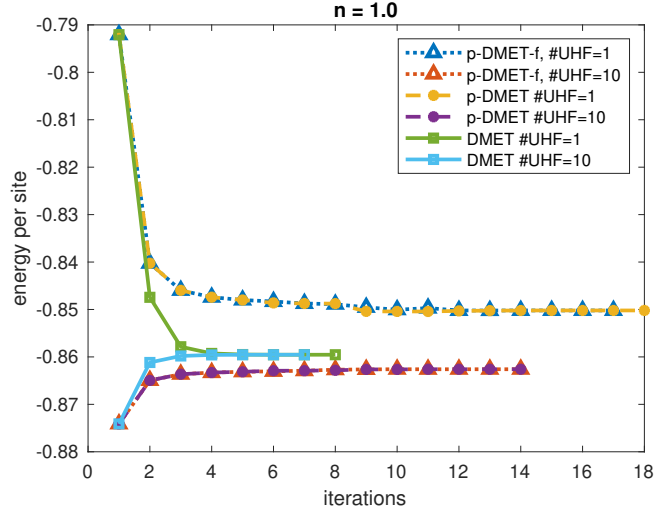}
  \caption{The convergence of the energy per site for p-DMET-f, p-DMET and DMET with different initial guesses. The system is a $6\times 6$ Hubbard model at $U=4.0$, $n=1.0$. }
  \label{fig:conv_half_fill}
\end{figure}

In the first example ($n=1.0$), DMET suggests that at convergence the system is in the AFM phase.  Therefore, we start from the PM phase, and break the spin symmetry of the initial density by alternately adding/subtracting a small number ($10^{-3}$) on odd/even sites to create slightly polarized spin-up and spin-down densities. Starting from this initial density, UHF converges within $20$ steps using DIIS. We input the initial 1-RDM for p-DMET/p-DMET-f after performing $1,5,10,20$ UHF iterations, respectively. For p-DMET-f, the gauge-fixing matrix is also obtained from the same 1-RDM. The convergence of the energies is reported in Fig. \ref{fig:conv_half_fill}. We find that the convergence curves of p-DMET and p-DMET-f are very similar and almost coincide with each other in all cases. Both the converged energy and the spin polarization from p-DMET/p-DMET-f depend on the initial guess of the 1-RDM. Table \ref{tab:polarization_half_fill} suggests that at convergence, UHF predicts an over-polarized spin configuration. However, starting from a significantly under-polarized 1-RDM obtained from one iteration of UHF, the converged solution of p-DMET underestimates the spin polarization (by $0.022$ relative
to the converged DMET result). With initial guesses obtained from an increased number of UHF iterations, both the energy and spin polarization obtained from p-DMET approach the results from DMET. After $5$-steps of UHF for the initial guess, p-DMET provides converged results in terms of energy and spin polarization. Remarkably, the solution of DMET is very robust with respect to the choice of the initial guess, even though neither DMET nor p-DMET/p-DMET-f guarantees a unique solution to the nonlinear fixed point problem \textit{a priori}.

\begin{table}[!htp]
  \centering
    \caption{Spin polarization for the $6\times 6$ Hubbard model at $U=4.0$, $n=1.0$ obtained from converged UHF, DMET and p-DMET calculations. \#UHF stands for the number of UHF steps to obtain the initial 1-RDM for the DMET calculation.} 
  \label{tab:polarization_half_fill}
  \begin{tabular}{c|c|c|c|c|c|c|c|c}
    \hline
     & initial
     & UHF  
     & \makecell{DMET \\(\#UHF=1)}
     & \makecell{DMET \\(\#UHF=5)}
     & \makecell{DMET \\(\#UHF=20)}
     & \makecell{p-DMET \\ (\#UHF=1)} 
     & \makecell{p-DMET \\(\#UHF=5)} 
     & \makecell{p-DMET \\(\#UHF=20)}\\
    \hline
    $m$ & 0.001 & 0.34876 & 0.30812 &  0.30812 & 0.30812 & 0.28578 & 0.30851 & 0.31000\\
    \hline
  \end{tabular}
\end{table}

\begin{figure}[!htp]
  \centering
  \includegraphics[scale=0.7]{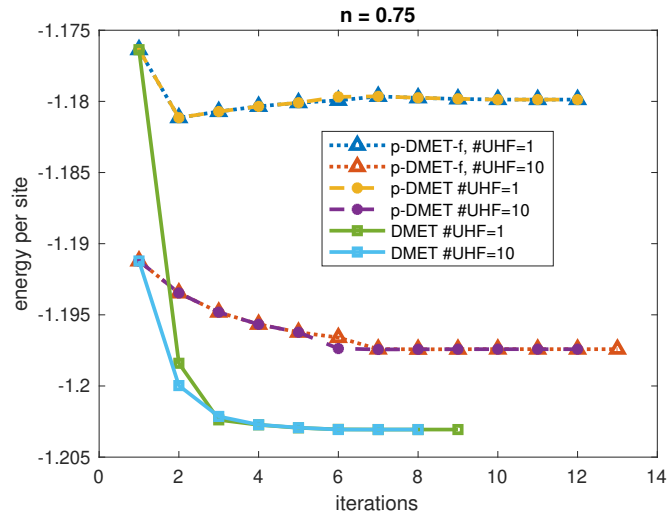}
  \caption{The convergence of the energy per site for p-DMET-f, p-DMET, and DMET with different initial guesses. $6\times 6$ Hubbard model at $U=4.0$, $n=0.722$.} 
  \label{fig:conv_away_half_fill}
\end{figure}

In the second example, we set $n=0.722$ ($N_e=26$). 
We start from an initial guess that exhibits AFM order, where the spin up component of the density is set to be $0.1444$ and $0.5778$ 
on alternate sites, and
the spin down component is arranged alternately in the opposite way with the same values. The convergence of the energy (Fig. \ref{fig:conv_away_half_fill}) is similar to that in the case of half-filling. For a spin-polarized initial 1-RDM obtained from one step of the UHF iteration, the converged solution of p-DMET remembers the initial guess and predicts an AFM phase with a small spin-polarization $0.04$. Both the energy and spin polarization improve quickly as the number of UHF iterations used to define the initial guess increases. Eventually p-DMET also predicts a PM phase. Again, no initial guess dependence is observed in DMET.


\begin{table}[!htp]
  \centering
    \caption{Spin polarization for the $6\times 6$ Hubbard model at $U=4.0$, $n=0.278$ obtained from converged UHF, DMET and p-DMET calculations. \#UHF stands for the number of UHF steps used to obtain the initial 1-RDM for the DMET and p-DMET calculations.} 
  \label{tab:polarization_away_half_fill}
  \begin{tabular}{c|c|c|c|c|c|c|c|c}
    \hline
      & initial & UHF  
      & \makecell{DMET \\(\#UHF=1)} 
      & \makecell{DMET \\ (\#UHF=5)} 
      & \makecell{DMET \\ (\#UHF=20)}  
      & \makecell{p-DMET \\(\#UHF=1)} 
      & \makecell{p-DMET \\ (\#UHF=5)} 
      & \makecell{p-DMET \\ (\#UHF=20)}\\
    \hline
    $m$ & 0.2167 & $2.166 \times 10^{-11}$  & $4.961\times 10^{-8}$  & $1.365 \times 10^{-7}$  & $5.440 \times 10^{-8}$ & $4.175 \times 10^{-2}$ & $2.719\times 10^{-2}$ & $4.865 \times 10^{-9}$ \\
    \hline
  \end{tabular}
\end{table}

In both examples above, UHF and DMET predict the same phase of matter. We find that self-consistent p-DMET can significantly reduce the error of physical observables starting from converged or unconverged UHF solutions, but there is some initial guess dependence. Therefore, when UHF and DMET predict different phases, p-DMET can reduce but not eliminate the (presumed) error from UHF. Hence the phase diagram in Fig. \ref{fig:phase} is similar to that of DMET, but softer near the boundary region.

\subsection{Effect of fragment size}\label{subsec:finite size convergence}
To understand the fragment size dependence of the physical observables in DMET and p-DMET, 
we carried out a
  number of 2D Hubbard model calculations (of $40 \times 40$ sites) at half-filling for different interaction strengths ($U = 2$, $4$, $6$ and $8$) and with different fragment sizes ($2\times2$, $2\times4$ and $4\times4$) using translational invariance. 
  The same cluster sizes were previously considered in Ref.~\cite{boxiao2016} where translationally invariant DMET is in the non-interacting bath (NIB) formulation only. We use the data from Ref.~\cite{boxiao2016} as reference. DMET in the interacting bath formulation, as used everywhere else in this work, is here denoted DMET (IB). Given a set of 2D clusters with $L_{A}$ sites, the DMET energy can be extrapolated to the
  thermodynamic limit (TDL) as a power series in $L^{-\frac{1}{2}}_{A}$ \cite{Zheng2017},
\begin{equation}
E(L) = E(\infty) + a_0 L^{-\frac{1}{2}}_{A} + b_0 \qty(L^{-\frac{1}{2}}_{A})^2 + \ldots
\end{equation}
We use the average of linear regression and a quadratic fit as the extrapolated result, and the error bar is defined to be the difference of the two fits. 

\begin{figure}[!htp]
  \includegraphics[scale=0.4]{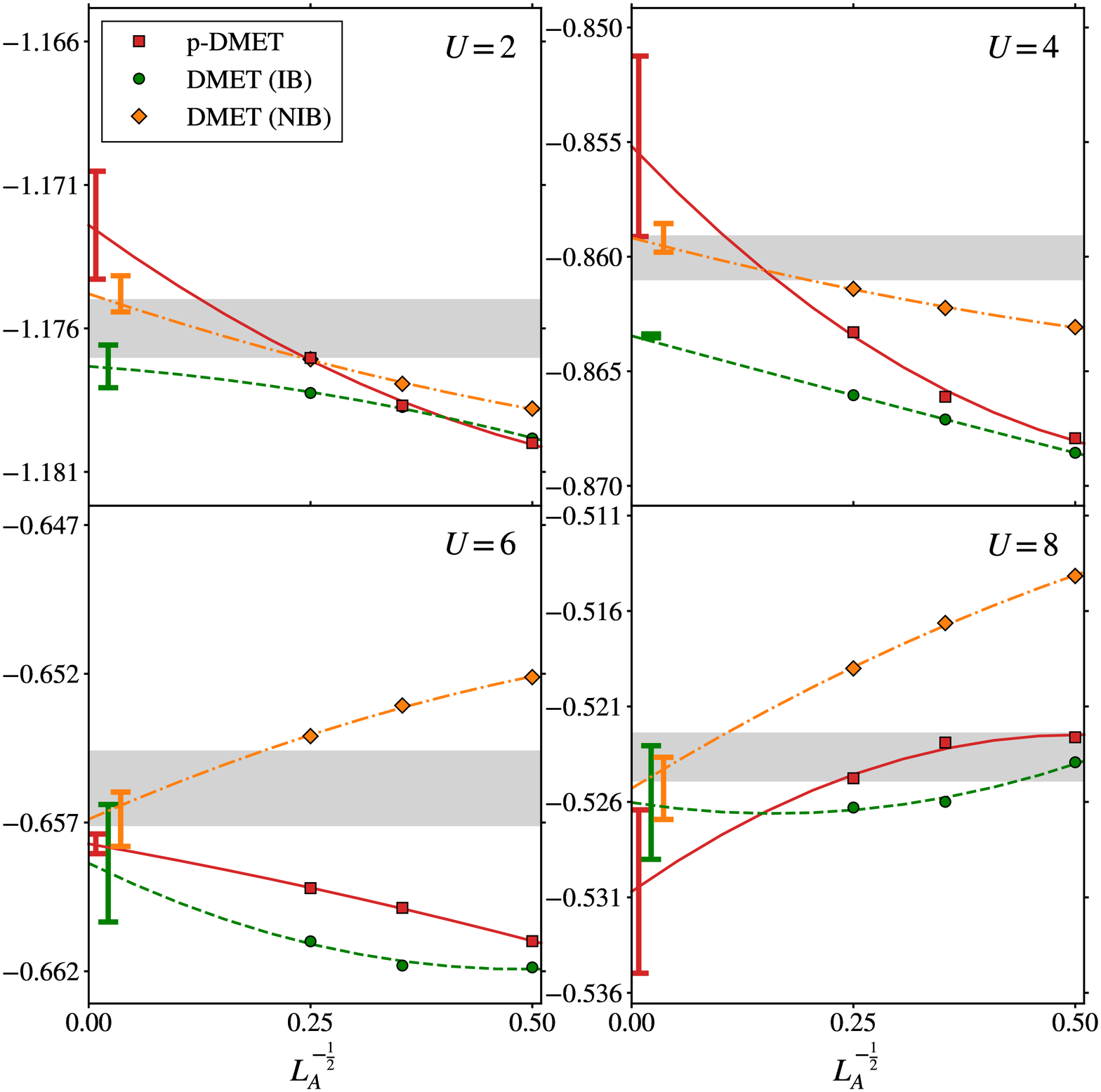}
  \caption{Energy extrapolation as a function of cluster size $L_{A}$ for p-DMET, DMET(IB) and DMET(NIB), where IB and NIB denote the interacting- and non-interacting bath formulations respectively. The fitting curve is an average of linear regression and quadratic fitting. The error bar is chosen to be the difference between the linear and quadratic extrapolated values. The shaded area is generated from the AFQMC, DMRG, DMET and DCA-DMET benchmark numbers
      in Ref.\cite{LeBlanc15} and Ref. \cite{Zheng2017}.}
  \label{fig:extrapolation}
\end{figure}

Fig.~\ref{fig:extrapolation} presents the calculated energy of three methods (p-DMET, DMET (IB), and DMET (NIB)) and the corresponding extrapolations. All three methods give reasonable extrapolated energy values, compared to the benchmark data (grey shaded region). For p-DMET, the largest error is about $5\times 10^{-3} $ at $U = 8$,
  while for DMET (IB), the largest error is about $3 \times 10^{-3}$ at $U = 4$. The behavior of DMET and p-DMET as a function of $L^{-\frac{1}{2}}_{A}$ is
  relatively similar, while that of DMET (NIB) is quite different. This reflects the energy influence of the different choice of impurity Hamiltonian. In fact, extrapolations
  using DMET (NIB) have smaller error bars in general, and they are slightly more accurate than those of DMET (IB) and p-DMET.  
 The results also indicate that after extrapolation, the performance of DMET and p-DMET is comparably reliable. 

\subsection{Efficiency} \label{subsec:efficiency}

To demonstrate the efficiency of p-DMET and p-DMET-f, we analyze two factors that affect the overall computational cost, i.e. the total number of iterations for self-consistent convergence and the (average) time cost per iteration. 

\begin{figure}[!htp]
    \centering
    \includegraphics[scale=0.4]{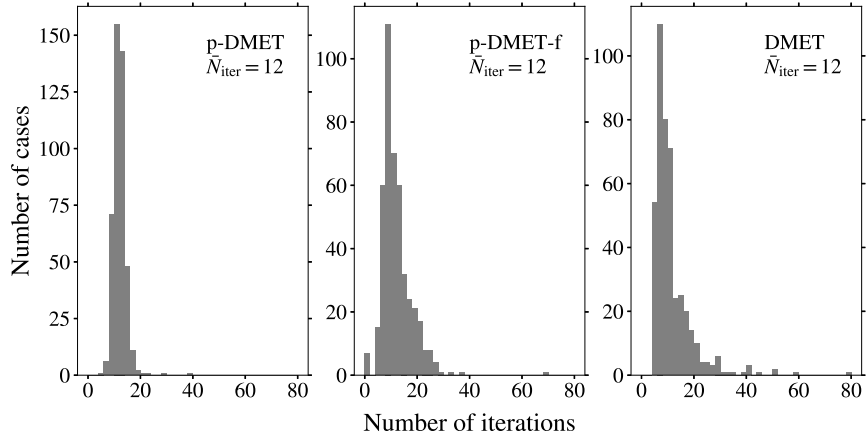}
    \caption{Distribution of the number of self-consistent iterations of p-DMET, p-DMET-f and DMET required
        to reach $10^{-6}$ in the energy across the phase diagram in Fig.~\ref{fig:phase}. All calculations used DIIS.}
    \label{fig:iter_num}
  \end{figure}

We first consider the number of iterations required for convergence. We extracted the number of iterations from the calculations
  for the preceding phase diagram (Fig. \ref{fig:phase}) and plot the distribution of the number of iterations in Fig. \ref{fig:iter_num}. As shown in the figure, all three methods (p-DMET, p-DMET-f and DMET) have a similar average convergence rate, with an average iteration number of 12 required to achieve an energy accuracy of $10^{-6}$. In most cases, the iteration number is less than $20$. We also remark that at least for systems studied in this work, the distribution in the case of p-DMET is slightly narrower with fewer outliers, and hence the self-consistent iteration is more stable.

Acquiring the similar average number of self-consistent iterations in p-DMET and DMET, we now discuss the total computational time. We performed a series of tests on lattices of size $2\times N$ ($N$ is the number of sites in the $y$ direction ranging from $8$ to $64$). We set $U=2.0$ and $n=1.0$ (half filling).  
We chose this quasi one-dimensional structure to ensure that the mean-field problem always has a positive energy gap, which is not the
case for arbitrary 2D lattices.
All the tests are performed on 36 Intel Broadwell vCPUs of the Google Cloud Platform (GCP).

We  measured the CPU time of DMET, p-DMET and p-DMET-f spent on single-particle type computations
and to solve the impurity problems (many-body computations).
When the fragment size is fixed, the cost to construct and solve the impurity problems (high-level computations) always scales linearly with respect to the global system size. In DMET, the
single-particle type computations (low-level computations) include the diagonalization of the mean-field Hamiltonian, and optimization of the
correlation potential; in p-DMET, the cost of the single-particle computations is mainly due to the  diagonalization to obtain the projected 1-RDM.

\begin{figure}[!htp]
  \centering
  \includegraphics[scale=0.8]{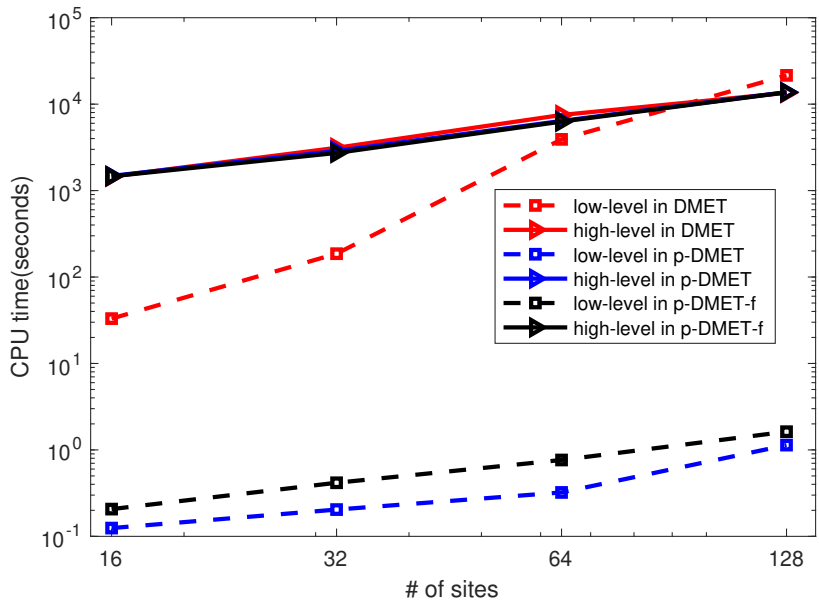}
  \caption{CPU time in DMET, p-DMET and p-DMET-f associated with (i) low-level density matrix computations (ii) high-level impurity problem computations.}
  \label{fig:cputime}
\end{figure}

In DMET, the single-particle computational cost significantly increases with  $N$ (Fig. \ref{fig:cputime}).
In each step of DMET, the correlation potential
optimization typically requires more than $100$ iterations to converge, and the number of iterations also grows with respect to the system size.
When the system becomes moderately large (number of sites larger than $64$), the cost at the mean-field level is much more
expensive than solving the $2 \times 2$ impurity problems. When the number of sites is $128$, the single particle computations in DMET take in total $\sim 20000$ s. 
On the other hand, the main single-particle cost in p-DMET is only the eigenvalue decomposition to obtain $D^{\text{ll}}$. Similarly, p-DMET-f only needs to perform single-particle type matrix multiplications and inversions. For the system with $128$ sites, the computational cost at the mean-field level of p-DMET is reduced to only $\sim 1$ s. This demonstrates that p-DMET/p-DMET-f provide a significant reduction in computational
cost relative to DMET for large, heterogeneous, systems.

\section{Conclusion}\label{sec:conclusion}

An important computational bottleneck in DMET calculations for large systems is the correlation potential optimization required
to achieve self-consistency. In this optimization,
each evaluation of the cost
function requires the diagonalization of a mean-field Hamiltonian, and each derivative evaluation amounts to
a response calculation. Thus, for a moderately sized system, the correlation potential optimization procedure can be expensive.  In this work, we viewed the self-consistent DMET as only one formulation of a more general fixed point problem to obtain the high-level global  density matrix $D^{\text{hl}}$.
From this  general perspective, we proposed the projected density matrix embedding method (p-DMET) as a simpler and
more efficient way to achieve self-consistency. We found that for the 2D Hubbard model, compared to the unrestricted Hartree-Fock (UHF) solution, p-DMET significantly improved the accuracy of the total energy and behaviour of the spin polarization across the entire parameter space. The phase diagrams predicted by p-DMET and DMET qualitatively agreed with each other, but the phase boundary obtained from p-DMET was softer.
Further investigation showed that this was because the self-consistent solution of p-DMET retained a weak dependence on the initial guess.
On the other hand, the cost associated with achieving self-consistency in p-DMET
was negligible compared to that needed to optimize the correlation potential in DMET.

There are a number of directions that should now be pursued. First, we would like to identify the root  of the initial
guess dependency of p-DMET.
Second, we plan to generalize p-DMET to superconducting systems, where the low-level theory requires the solution of the Bogoliubov-de Gennes (BdG) equations.
Third, a remaining numerical issue in DMET associated with self-consistency is the appearance of vanishing gaps in the DMET mean-field Hamiltonian
during the optimization. This should then be treated using a zero temperature limit of a finite temperature formulation of DMET. We will report these works in future publications.

\section*{Acknowledgment}

This work was partially supported by the Air Force Office
of Scientific Research under award number FA9550-18-1-0095 (XW, ZHC, GKLC and LL),  
by the Department of Energy under
Grant No. DE-SC0017867 (XW and LL),  by the Department of Energy
CAMERA program (YT and LL), and by the National Science Foundation
Graduate Research Fellowship Program under grant DGE-1106400 (ML).   We
thank Berkeley Research Computing (BRC), Google Cloud Platform (GCP) and Caltech High Performance Computing Center (Caltech HPC) for computing resources.

\appendix
\section{Obtaining bath orbitals}\label{sec:bath}
Let $D^{\text{ll}}$ be a mean-field density matrix. Without loss of generality, let us introduce a rotation matrix $R_x$ for each fragment
$x$ 
such that the rotated density matrix $D = R_x D^{\text{ll}}R_x^\dagger$ can be written as
\begin{equation}
  \label{eq:7}
  D = CC^\dagger =
  \begin{bmatrix}
    D_{11} & D_{12}\\
    D_{21} & D_{22}
  \end{bmatrix},
\end{equation}
where $D_{11}$ corresponds the fragment $x$ only. We divide the matrix $C$, which contains the occupied orbitals, into two matrices
\begin{equation}
  \label{eq:14}
  C =
  \begin{bmatrix}
    C_A \\ C_B
  \end{bmatrix} .
\end{equation}
$C_A$ is the rectangular matrix ($L_A\times N_e$)
with rows in the fragment (not fragment orbitals), and $C_B$ is the rectangular matrix ($(L-L_A)\times N_e$) with rows in the environment (not bath orbitals). By diagonalizing the matrix $D_{11}$, we have 
\begin{equation}
  \label{eq:8}
  D_{11} = U_A\Sigma_A^2U_A^\dagger ,
\end{equation}
where $\Sigma_A$ is a diagonal matrix containing all $L_A$ singular values of $C_A$. $U_A$ is a $L_A\times L_A$ unitary matrix. We can simply choose $V_A = I_{L_A}$ such that the above diagonalization corresponds the SVD on $C_A$:
\begin{equation}
  \label{eq:9}
  C_A = U_A\Sigma_AV_A^\dagger.
\end{equation}
By the orthogonality of the orbitals, $C_B$ can be uniquely expressed as
\begin{equation}
  \label{eq:13}
  C_B=U_B\Sigma_BV_A^\dagger ,
\end{equation}
where $U_B$ is to be determined. 
Thus, the bath-fragment density matrix can be written as
\begin{equation}
  \label{eq:15}
  D_{21}=C_BC_A^\dagger=U_B\Sigma_B\Sigma_AU_A^\dagger.
\end{equation}
The unitary matrix $U_B$ can be calculated by normalizing all the columns of the matrix $D_{21}U_A$, since $$U_B\Sigma_B\Sigma_A=D_{21}U_A.$$ The diagonal elements of $\Sigma_A\Sigma_B$ are the corresponding norms of the columns. As a result, $\Sigma_B$ is also obtained with the known $\Sigma_A$ in \eqref{eq:8} Thus, we have the fragment orbitals and bath orbitals, respectively
\begin{equation}
  \label{eq:16}
  C_f =
  \begin{bmatrix}
    U_A\\0
  \end{bmatrix}
  , C_b =
  \begin{bmatrix}
    0\\ U_B
  \end{bmatrix} .
\end{equation}
The fragment orbitals $C_f$ and bath orbitals $C_b$ are then combined into the DMET impurity problem orbitals 
\begin{equation}
  C_x = \begin{bmatrix}C_f & C_b \end{bmatrix} .
  \label{eqn:Cx}
\end{equation}
Since the core orbitals do not explicitly appear in the DMET formulation, they do not have to be computed explicitly. The contribution of the core orbitals is reflected by the core density matrix. The core density matrix is expressed as
\begin{equation}
    D^{\text{core}} = D_{22} - U_B\Sigma_B^2U_B^\dagger .
\end{equation}
The above implementation is equivalent to the diagonalization of the density matrix $D_{22}$. In fact,
\begin{equation}
  \label{eq:11}
  C_BC_B^\dagger=
  \begin{bmatrix}
    U_B & U_C
  \end{bmatrix}
  \begin{bmatrix}
    I_{L_A}-\Sigma_B^2 & 0\\
    0 & I_{N_e-L_A}
  \end{bmatrix}
  \begin{bmatrix}
    U_B^\dagger\\
    U_C^\dagger\\
  \end{bmatrix} .
\end{equation}
During the above construction of the bath orbitals, we only need to perform an SVD on the matrix $C_AC_A^\dagger$ instead of diagonalizing the matrix $C_BC_B^\dagger$ or performing SVD on $C_AC_B^\dagger$. This process reduces the computational cost of bath orbitals construction from $O(N_e^3)$ or $O(N_eL_A^2)$ to $O(L_A^3)$ when $N_e\gg L_A$.

\bibliography{pdmet_updated}

\end{document}